\documentclass[10pt, conference, compsocconf]{IEEEtran}
\ifCLASSINFOpdf
\else
\fi
\usepackage{graphicx}
\usepackage{caption}
\usepackage{epstopdf}
\usepackage{todonotes}
 \usepackage{url}

\begin{document}
%
\title{A new analysis of Work Stealing with latency}



%

\author{\IEEEauthorblockN{
		Mohammed Khatiri\IEEEauthorrefmark{1}\IEEEauthorrefmark{2},
		Denis Trystram\IEEEauthorrefmark{1},
		Frederic Wagner\IEEEauthorrefmark{1},
	}\\
	\IEEEauthorblockA{\IEEEauthorrefmark{1}Univ. Grenoble Alpes, CNRS, Inria, LIG,
		F-38000, Grenoble, France\\
	\IEEEauthorblockA{\IEEEauthorrefmark{2}Univ. Mohamed First, Faculty of Sciences, LaRI,
		60000 Oujda, Morocco\\
		Email: \{mohammed.khatiri,denis.trystram,frederic.wagner\}@imag.fr}
		}
}


%


\maketitle

\begin{abstract}
\noindent
We study in this paper the impact of communication latency on
the classical \emph{Work Stealing} load balancing algorithm.
Our approach considers existing performance models and
the underlying algorithms. We introduce a latency parameter
in the model
and study its overall impact by careful observations of
simulation results.
Using this method we are able to derive a new expression
of the expected running time of divisible load applications.
This expression enables us to predict under which conditions a given
run will yield acceptable performance. 
For instance, we can easily
calibrate the maximal number of processors one should use for a given
work/platform combination.
We also consider the impact of several algorithmic variants like
simultaneous transfers of work or thresholds for avoiding useless transfers.
All our results are validated through simulation on a wide range
of parameters.
\end{abstract}


%
\IEEEpeerreviewmaketitle

\section{Introduction}
\label{sec:intro}

\subsection{Why does latency matter?}
Distributed memory clusters consist in independent processing elements (called \emph{nodes}) with private local memories linked by an interconnection network. 
The architecture of the cluster is usually organized in several levels of hierarchies (where for instance the nodes are linked by fat trees~\cite{Hoefler2011}) or by interconnection topologies (multi-dimensional torus, dragon fly, etc.~\cite{Cheriere2016,Bhatele2008})
and their interconnection highly influences the performances of the applications deployed on such machines~\cite{MCMCA}.
Recent studies show that communication issues are crucial.
However, there are only few works dealing with optimized allocation strategies and the relationships with the allocation and scheduling process is most often ignored.
In practice, the impact of scheduling may be huge since the whole execution can be highly affected by a large communication latency of interconnection networks~\cite{MultiCoreClusterArchitectur2015}.

Scheduling is the process which aims at determining where and when to execute the tasks of the target parallel application. 
The applications are represented as directed acyclic graphs where the vertices are the basic operations and the arcs are the dependencies between the tasks~\cite{CosnardTrystram92}.
Scheduling is a crucial problem which has been extensively studied under many variants for the successive generations of parallel and distributed systems.
The most common studied objective is to minimize the maximum completion time of the tasks (called \emph{Makespan} and denoted by $C_{\max}$) and the underlying context is usually to consider centralized algorithms.
This assumption is not always realistic, especially if we consider distributed memory allocations and an on-line setting.

Work Stealing is an efficient distributed scheduling mechanism targeting medium range parallelism of multi-cores for fine-grain tasks. 
Its principle is briefly recalled as follows:
each processor manages its own (local) list of tasks. 
When a processor becomes idle it randomly chooses another processor and steals some work (if possible). 
Work Stealing has been implemented successfully in several languages and parallel libraries including Cilk~\cite{Leiserson1998,Leiserson2009}, TBB (Threading Building Blocks)~\cite{Robison2008}, the PGAS language~\cite{Dinan2009,Seung-Jai2011} and 
the KAAPI run-time system~\cite{Kaapi2007}. 
Its analysis is probabilistic since the algorithm itself is randomized.

We are interested in this work in studying how different latencies impact Work Stealing.
We show how classical expressions
of the expected Makespan require indeed some adjustments in the context of modern distributed systems. 
We also exhibit interesting behavior from different variants of the algorithm.

\subsection{Related works}
\label{subsec:relatedworks}

We briefly review the most relevant theoretically-oriented works.
Work Stealing has been analyzed originally by Blumofe and Leiserson in~\cite{Blumofe1999}.
They show that the expected Makespan of a series-parallel precedence graph
with $W$ unit tasks on $p$ processors is bounded
by $E(C_{\max}) \leq \frac{W}{p}+\mathcal{O}(D)$ where $D$ is the length of the
critical path of the graph (its depth).
This analysis has been improved in Arora \textit{et al.}~\cite{Arora2001} using potential functions. 
The case of varying processor speeds has been analyzed by Bender and Rabin in~\cite{Bender2002}
where the authors introduced a new policy called \textit{high utilization scheduler} that extends the homogeneous case. 
The specific case of tree-shaped computations with a more accurate model has been analyzed in~\cite{Sanders1999}. 
However, in all these previous analyses, the precedence graph is constrained to have only one source and an out-degree of at most 2 which does not easily model
basic case of independent tasks. 
Simulating independent tasks with a binary precedences tree
gives a bound of $\frac{W}{p}+\mathcal{O}(\log_2 (W))$ since a complete binary tree of $W$ vertices has a depth $D \leq \log_2 (W)$. 
However, with this approach, the structure of the binary tree dictates which tasks are stolen.

In complement,~\cite{Gast2010} provided a theoretical analysis based on a Markovian model using mean field theory. They targeted the expectation of the average response time
and showed that the system converges to a deterministic Ordinary Differential Equation.

The best existing bounds achieve a bound of the same order of the ratio
$\frac{W}{p}$ with a better constant of the term in $\log_2(W)$ and where
the processors are free to choose which tasks to steal. 
Note that there exist other results that study the steady state performance
of Work Stealing when the work generation is random including
Berenbrink \textit{et al.}~\cite{Berenbrink2003}, Mitzenmacher~\cite{Mitzenmache1998},
Lueling and Monien~\cite{Lueling1993} and Rudolph \textit{et al.}~\cite{Rudolph1991}.
More recently Tchiboukjian \textit{et al.} provided the best bound known at this time: $\frac{W}{p}+c.(\log_2 W)+\Theta(1)$
where $c$ is a small positive constant which can be determined precisely\cite{Denis2013}. 

In all these theoretical analyses, communications are not directly addressed 
(or at least are taken implicitly into account by the underlying model).
Besides theoretical works there exist more practical studies implementing Work Stealing libraries where some attempts were provided for taking into account communications:

SLAW is a task-based library introduced in~\cite{Gue2010}, combining work-first
and help-first scheduling policies focused on locality awareness 
in PGAS (Partitioned Global Address Space) languages like UPC (Unified Parallel C). 
It has been extended in HotSLAW, which provides a high level API that abstracts concurrent task management~\cite{Seung-Jai2011}.
\cite{Shigang2013} proposes an asynchronous Work Stealing (AsynchWS) strategy which exploits opportunities to overlap communication with
local tasks allowing to hide high communication overheads in distributed memory systems. 
The principle is based on a hierarchical victim selection, also based on PGAS.
Perarnau and Sato presented in~\cite{swann2014} an experimental evaluation of Work Stealing on the scale of ten thousand compute nodes where the communication depends on the distance between the nodes.
They investigated in detail the impact of the communication on the performance. In particular, the physical distance between remote nodes is taken into account.

Mullet \textit{et al.} studied in~\cite{Muller2106} 
Latency-Hiding, a new Work Stealing algorithm that hides the overhead
caused by some operations, such as waiting for a request from a client or waiting
for a response from a remote machine.
The authors refer to this delay as \emph{latency} which is slightly different that the more general concept that we consider. 
Agrawal \textit{et al.} proposed an analysis~\cite{Agrawal} showing the optimality for task graphs with bounded degrees
and developed a library in $Cilk++$ called \textit{Nabbit} for executing tasks
with arbitrary dependencies, with reasonable size blocks (granularity).

\subsection{Contributions and content of this paper}
\label{subsec:contributions}

The main contribution of this work is to realize an in-depth study of
the different impacts of latency on Work Stealing.
Our main goal is to gather observations on different variants of Work Stealing
under varying latencies in order to gain a good understanding of
their respective behaviors.

More precisely, we provide the following contributions:

\begin{itemize}
\item We derive a new expression of the expected Makespan which takes into account the latency $\lambda$:
\begin{center}
	$E(C_{\max}) = \frac{W}{p} + 2\lambda.c'.\log_2(\frac{W}{2\lambda})$
\end{center}
\item We study several stealing policies which all achieve the previous bound.
  \begin{itemize}
    \item We introduce a steal threshold in the policy. The goal here is to prohibit useless steals.
    \item We study simultaneous transfers on the same processor.
  \end{itemize}

\item We develop a python discrete event simulator for running adequate experiments.
This simulator uses the Work Stealing algorithm to schedule an amount of work $W$ 
in a distributed platform composed of $p$ identical processors.
The code is available on github\footnote{\url{https://github.com/wagnerf42/ws-simulator}}
along logs of presented experiments.
The simulator is generic enough to be used in different contexts of online
scheduling and interfaces with standard trace analysis tools (Paje file format~\cite{Paje2000}).
\end{itemize}

We start by recalling the classical Work Stealing mechanism in Section~\ref{sec:WSMechanisms}.
Then, we discuss several ways to take communications into account.
This section details in particular several variants of Work Stealing (namely, with simultaneous responses and a steal threshold).
Section~\ref{sec:expectedakespan} studies how to derive a new expression of the expected Makespan with latency. This is done via the careful study of the evolution of the number of steals for various latencies.
Section~\ref{sec:experiemnts} presents the experimental campaign for a deeper understanding of the role of the latency.
More specifically, we report the analysis assessing the quality of the Makespan expression and
analyze the impact of simultaneous responses.
Finally, Section~\ref{sec:conclusion} concludes this work and opens up some perspectives.

\section{Work Stealing Mechanisms}
\label{sec:WSMechanisms}

In this paper we are interested in studying the Work Stealing algorithm in a platform where communications are significant. 
Work Stealing is a decentralized list scheduling algorithm where each processor $P_i$ maintains its own local queue $Q_i$ of tasks to execute. 
$P_i$ uses $Q_i$ to get and execute tasks while $Q_i$ is not empty. 
When $Q_i$ becomes empty $P_i$ chooses another processor $P_j$ randomly and sends to it a steal request. The answer of $P_j$ can be to transfer some of its work or a fail response.

\begin{figure}[h!]
	\hspace{-15pt}
	\includegraphics[width=1.1\linewidth]{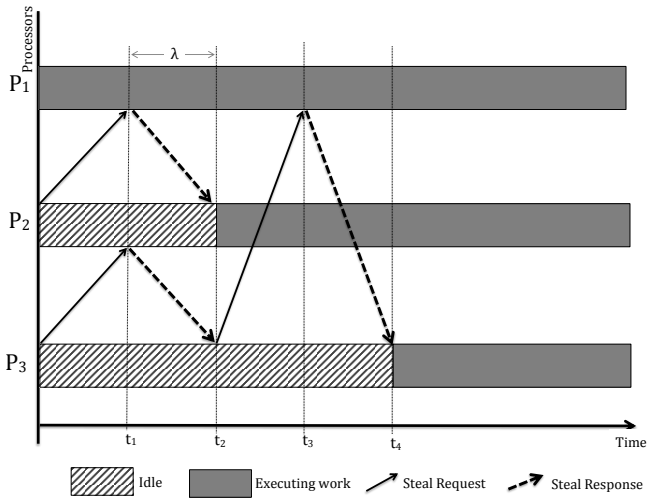}
	\caption{example of a Work Stealing execution} 
\label{fig:workStealing}
\end{figure}

Fig~\ref{fig:workStealing} depicts a simple execution of Work Stealing with latency.
We consider in this example that $\lambda=5$ and the total work is $W=100$.
At $t=0$ we have works on $P_1$, $P_2$, $P_3$ of respectively $(100, 0, 0)$. 
At this time $P2$ (resp. $P_3$) sends a work request to $P_1$ (resp. $P_2$).
Both $P_1$ and $P_2$ receive the work request at $t_1 = t_0 + \lambda = 5$. 
$P_1$ responds to $P_2$ by sending half of its work while $P_2$ responds to $P_3$ by a fail request and the remaining work become
$(48, 0, 0)$. At $t_2 = t_0 + 2\lambda=10$ $P_2$ receives its incoming work and the works are updated to
$(43, 47, 0)$. $P_3$ randomly chooses $P_1$ and sends again a steal request.
At $t_3$ $P_1$ receives the steal request from $P_3$ and responds by half of its work. 
At $t_4$ $P_3$ receives the response and works become $(14, 37, 19)$.       

Many algorithms and implementation variants of the Work Stealing algorithm exist in the literature.
We clarify in this section the most important mechanisms considered in our analysis. 
In particular, we present how the work is managed and how the communications between processors is handled. 
We also describe different policies for steal answers.


\subsection{Task model}

The computational load (called \emph{work}) is usually stored as tasks and their dependencies. 
Another possibility is to consider the work as a \emph{divisible load} where the initial amount is represented by a single big task. 
Then, during execution, each task can be divided on request into two subtasks containing each half of its work. 
For instance when a steal request occurs in a busy processor it sends a positive response
in a form of a new task containing half of the local work and updates accordingly its current content.
Many theoretical studies on Work Stealing use this divisible load model since it simplifies the analysis~\cite{Denis2013}. 

Note that this model fits the behavior of real world applications rather well 
since divisible load is indeed very close to fine-grain independent tasks.
There also exist adaptive applications that are able to create tasks on demand~\cite{JeanLouisRoch2006} which may be considered as a divisible load.

\subsection{Communication issues}

It is very common nowadays to handle communications through a thread dedicated to sending and receiving operations. 
This technique allows for a good overlap of communications and computations.
Moreover, several communications can take place simultaneously with no extra overhead when the communication costs are dominated by the latency.

According to the architecture of modern processors and the fine-grain model of tasks, any communication can be modeled by a constant delay (the inter-node latency, denoted by $\lambda$). 
Our objective is to study the behavior of the classical Work Stealing algorithm under a simple model. 
Using a latency-based network model enables us to have a good understanding of observed behaviors and an easier link with the theoretical analyses.

\subsection{Simultaneous responses}
There exist in the literature two main variants for handling steal responses, namely, the single and simultaneous responses. 
We consider here both techniques and provide a performance comparison between them in Section~\ref{Sec:simultaneousresponse}.

\begin{figure}[h]
	\hspace{-15pt}
	\includegraphics[width=1.1\linewidth]{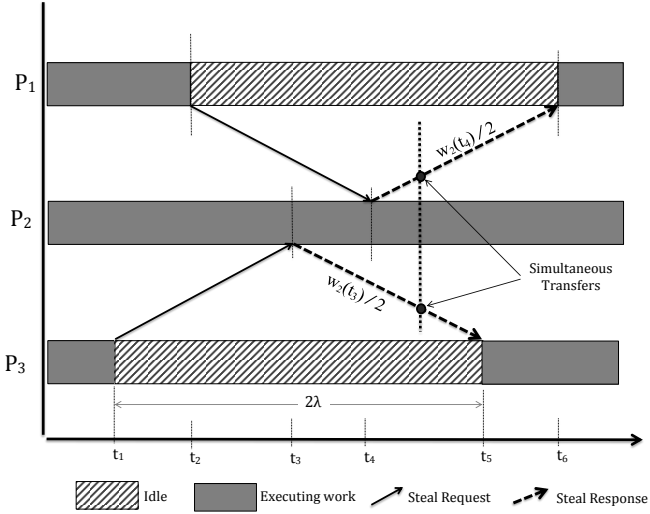}
	\caption{distribution of work in case of simultaneous responses} 
\label{fig:simultaneousdistribution}
\end{figure}

\begin{itemize}
	\item
	\textbf{Single work transfer SWT} 
	is a variant where the processor can send some work to at most one processor at a time. 
	While the processor sends work to a thief it replies by a fail response to any other steal request. 
        Using this variant the \emph{steal request} may fail in the two following cases: 
	when the victim does not have enough work or when it is already sending some work to another thief.
	\item
	\textbf{Multiple work transfers MWT }
	is the variant where each processor can respond and send work to several processors simultaneously. 
	The received requests are handled sequentially.
	The processor always answers by sending half of its work. 
	In case of simultaneous requests it serializes them and answers in the same way.
    Fig~\ref{fig:simultaneousdistribution} gives an example of such simultaneous work transfers. In this figure $W_i(t)$ denotes work on $P_i$ at time $t$.
\end{itemize}

\subsection{Steal Threshold}

The main goal of Work Stealing is to share work between processors in order to balance the load and speed-up the execution. 
In some cases however it might be beneficial to keep work local and answer negatively to some steal requests.

\begin{figure}[h!]
	\vspace{-35pt}
	\hspace{-15pt}
	\includegraphics[width=1.15\linewidth]{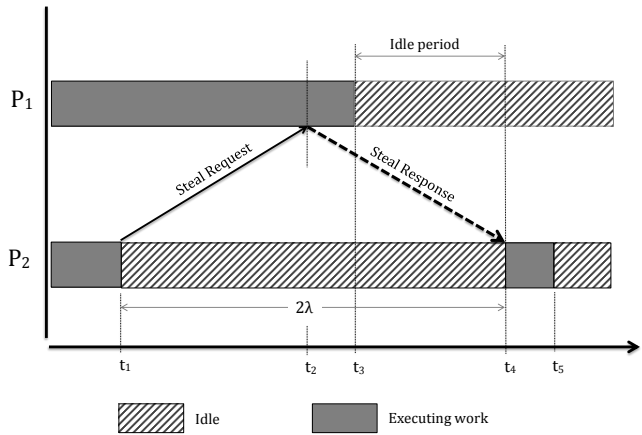}
	\caption{example of creating artificial idle times} 
\label{fig:stealthreshold}
\end{figure}

Fig~\ref{fig:stealthreshold} shows an example of this case on two processors. 
At time $t_1$ processor $P_2$ sends a steal request to $P_1$. 
At $t_2$ $P_1$ receives this request and answers by sending half of its local work, which is less than the communication duration. 
At $t_3$ $P_1$ finishes its remaining work and becomes idle. 
Then, both processors are idle in the time period between $t_3$ and $t_4$.
This is clearly a waste of resources since the whole platform is idle while it remains some work to execute. 
Moreover, such a behavior can be chained several times. 
This effect is not purely theoretical as it has been observed during our initial experiments campaign.  
It is possible to prevent this from happening by adding a threshold on steal operations. 
We introduce a \emph{steal threshold} which prohibits steals if the remaining local work becomes too small.
We set this threshold to $2\lambda$ in order to avoid useless transfers.

\section{A new expression for the expected Makespan}
\label{sec:expectedakespan}
The objective of this section is to determine an expression for the expected Makespan of Work Stealing with latency.
We show here how to derive such an expression.
The methodology leads to an accurate expression which remains relatively simple.
It is also stable for all variants of the Work Stealing model discussed in the previous sections.

The experiments performed in this paper use our discrete events simulator.
We recall that simulation parameters are the work $W$, a number $p$ of processors and latency $\lambda$.

\subsection{Analysis}

Assuming the existing simplified model with no communications the most accurate expression of the expected Makespan is:
\begin{center}
    $E(C_{\max}) = \frac{W}{p} + c.\log_2(W) + \Theta(1)$
\end{center}
where $c$ is a small positive constant~\cite{Denis2013}.
The expected number of steal operations is given by the following expression:
\begin{center}$E(R) = c.\log_2(W). p$\end{center}
\medskip

\begin{figure}[h]
	\centering
	\includegraphics[width=1\linewidth]{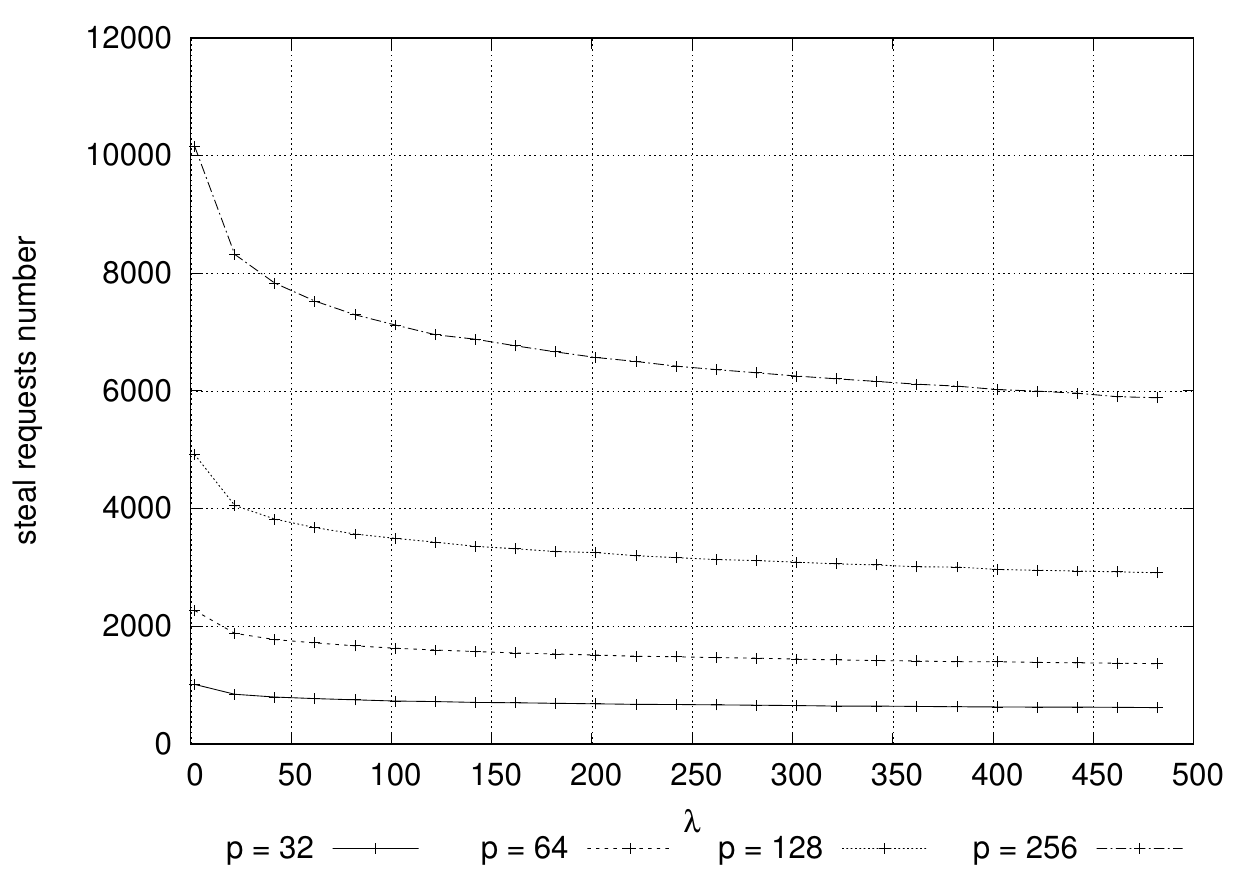}
	\caption{number of steal requests  according to latency for different numbers of processors ($W=10^8$)}
\label{fig:stealLatency}
\end{figure}
The natural idea for taking into account the latency of communications is to penalize the number of steals proportionally to the latency.
However this simple idea does not fit with simulated results.
Fig~\ref{fig:stealLatency} depicts a simulation of the number of steals with regard to the latency $\lambda$ on a typical example.
We can see that it evidences a non-linear behavior.

\subsection{Proposed expression}

Fitting the previous curves leads to the following expression:
\begin{center}
    $E(C_{\max}) = \frac{W}{p} + 2\lambda.c'.\log_2(\frac{W}{2\lambda})$
\end{center}

It is possible to understand this formula by normalizing $W$ by $2\lambda$, using the classic formula
and re-multiplying time estimations by $2\lambda$.
This view corresponds to an execution of macro-s-eps-converted-to.pdf of duration $2\lambda$.
In practice communications happen any time and could happen inside such s-eps-converted-to.pdf.
However, this macro-step view is leading to an estimation of $C_{\max}$ close to the observed results.

The constant $c'$ has been estimated to a value of $1.8$ by fitting data over a large set of parameters.
This gives us the following formula for Makespan estimation:

\begin{center}
    $E(C_{\max}) = \frac{W}{p} + 3.6\lambda.\log_2(\frac{W}{2\lambda})$
\end{center}

\section{Experiments Analysis}
\label{sec:experiemnts}
\subsection{Experiments}

The objective of this section is to study the impact of latency on Work Stealing in an experimental setup. 
We start by assessing the validity of our Makespan Formula and showing the latency intervals exhibiting an acceptable Makespan. 
Finally we conclude this section by studying the impact of simultaneous responses.


Before starting to analyze the experiments let us describe our experimental parameters.
On the side of processors we consider constant speed processors. 
Since all of them have the same speed, the work can be described as a time unit, and the same holds for the latency. You can consider our time unit as milliseconds although ultimately only the ratio between $\frac{W}{p}$ and $\lambda$ really matter. 
Similar results would be observed by multiplying $\frac{W}{p}$ and $\lambda$ by the same constant.\\
For most of our tests we take different parameters with $W$ between $10^5$ and $10^8$,
$p$ between 32 and 256 and $\lambda$ between 2 and 500. Each experiment has been reproduced over 1000 runs.

\subsection{Expected Time}

\begin{figure}[h!]
	\centering
	\includegraphics[width=\linewidth]{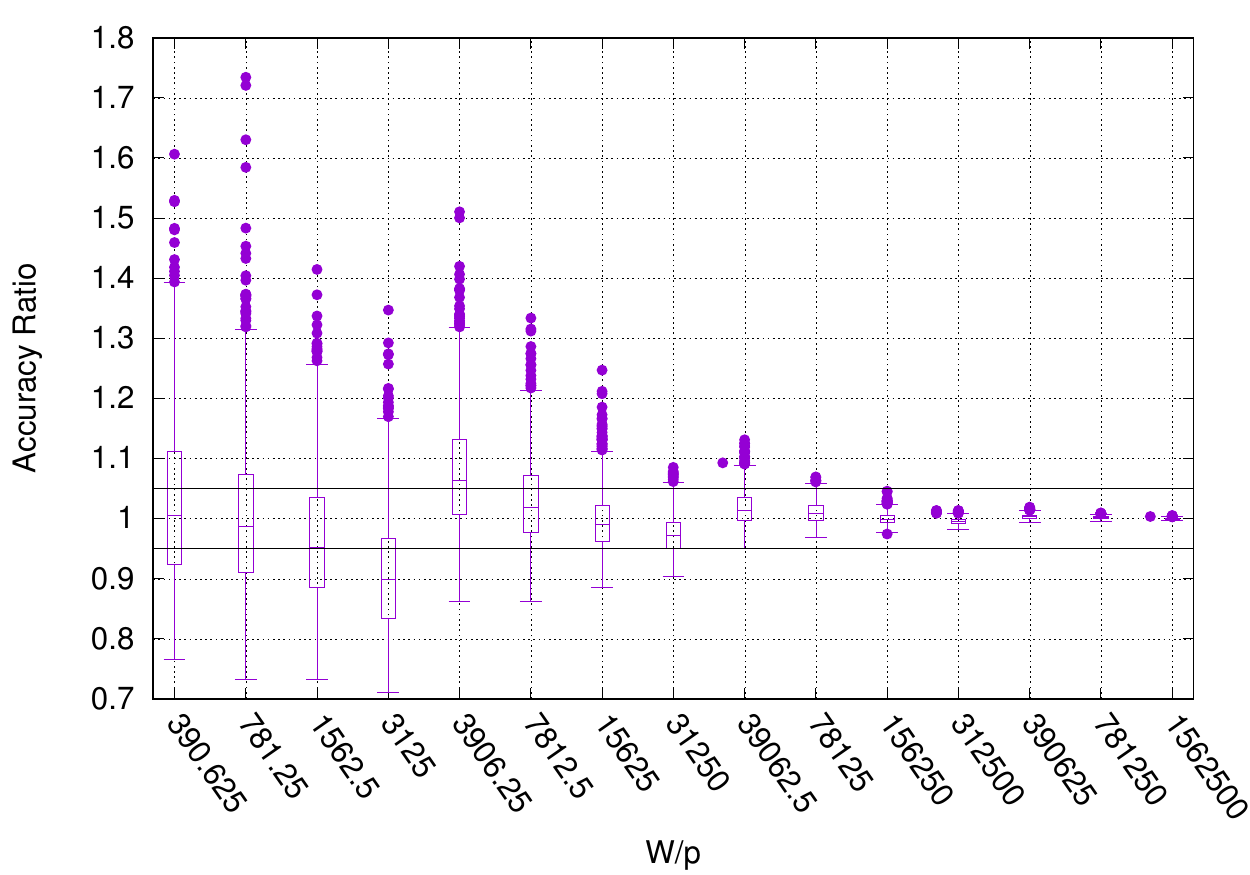}
	\caption{accuracy ratio of our model according to the $\frac{W}{p}$ ratio  ($\lambda = 262$)}
\label{fig:accuracy}
\end{figure}
We start by examining the efficiency of the new Makespan expression.
We therefore consider the ratio between the simulation time and the predicted time. 
We define this ratio as the accuracy ratio and we study it under different parameters.\\
All three $W$, $p$ and $\lambda$ parameters impact our estimation.
Fig~\ref{fig:accuracy} plots the accuracy ratio according to the ratio $\frac{W}{p}$ for a medium range latency of 262. 
The~x-axis is $\frac{W}{p}$ for all our $W$ and $p$ intervals and the y-axis shows the accuracy ratio. 
We use here a BoxPlot graphical method to present the results. BoxPlots give a good overview and a numerical summary of a data set. The “interquartile range” in the middle part of the plot represents the middle quartiles where 50\% of the results are presented. The line inside the box presents the median. The whiskers on either side of the IQR represent the lowest and highest quartiles of the data.  The ends of the whiskers represent the maximum and minimum of the data, and the individual dots beyond the whiskers represent outliers.

We observe immediately that the model accuracy is increasing strongly with $\frac{W}{p}$. 
The reason is that the $\frac{W}{p}$ Makespan lower bound $\frac{W}{p}$
grows linearly with $W$ while the overhead $3.6\log_{2}(\frac{W}{2\lambda})\lambda$ increases logarithmically with $W$. 
We can also see that as soon as  $\frac{W}{p}$ reaches 60x the latency ($\frac{W}{p} \geq 15625$) more than 50\% of runs are full under 5\% of error.  
Similar observation have been observed with all values of $\lambda$ used.

\begin{figure}[h!]
	\centering
	\includegraphics[width=\linewidth]{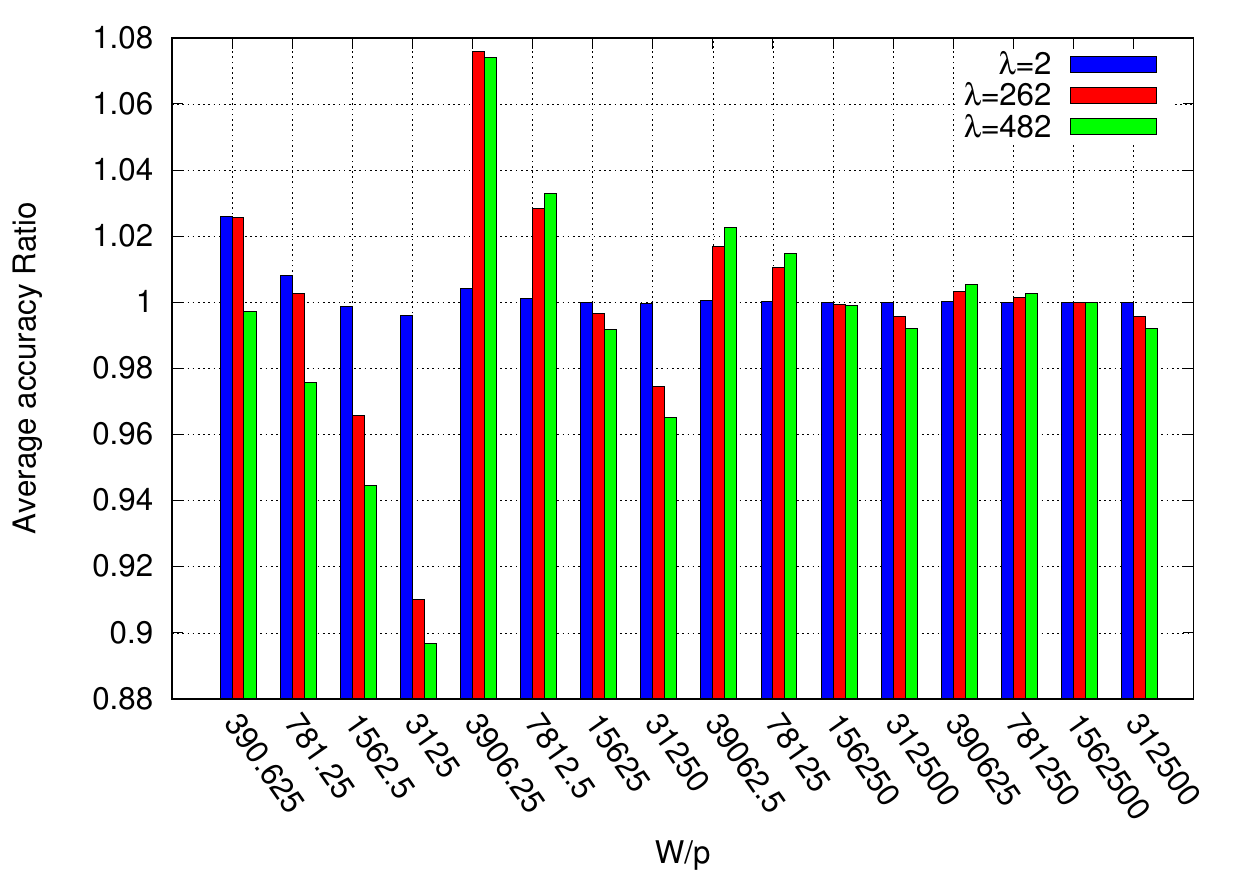}
	\caption{average accuracy according to $\frac{W}{p}$ for varying values of $\lambda$ } 
\label{fig:histogram}
\end{figure}

In all our results, the maximum error for the middle quartiles is reached for $\frac{W}{p}=3125$ and $\lambda=482$ with a maximal error
on average Makespan of less than 11\% (Fig~\ref{fig:histogram}).
Since no runs exceed this value we conclude positively on the quality of our prediction.

\subsection{Acceptable latency}
One of the first uses of the analytical expression for time predictions is to be able to predict when a given $\frac{W}{p}$ and $\lambda$ configuration will yield acceptable performances. 
Using the Makespan expression we observe that two parameters dominate: The $\frac{W}{p}$ ratio in the first term and $\lambda$ which impacts the second term of the formula representing the overhead due to communication delays. 

As stated before $\frac{W}{p}$ is a good lower bound on the best possible Makespan. A Makespan $C_{\max}$ is acceptable if the ratio $\frac{C_{\max}}{C_{\max}^*}$ is close to 1, where $C_{\max}^*$ is the best possible Makespan.
In our analysis, we consider a Makespan $C_{\max}$ as acceptable if $\frac{C_{\max}}{\frac{W}{p}} \leq 1.1$ (overhead less than 10\%).
We study here which configurations allow us to obtain such an acceptable Makespan.
Using the time estimation Formula we derive the equation below linking $W$, $\lambda$ and $p$
in order to get an acceptable Makespan.

\begin{center}
	$\frac{W}{p} + 3.6\log_{2}(\frac{W}{2\lambda})\lambda = 1.1\frac{W}{p} $
\end{center}
\begin{figure}[h!]
	\centering
	\includegraphics[width=\linewidth]{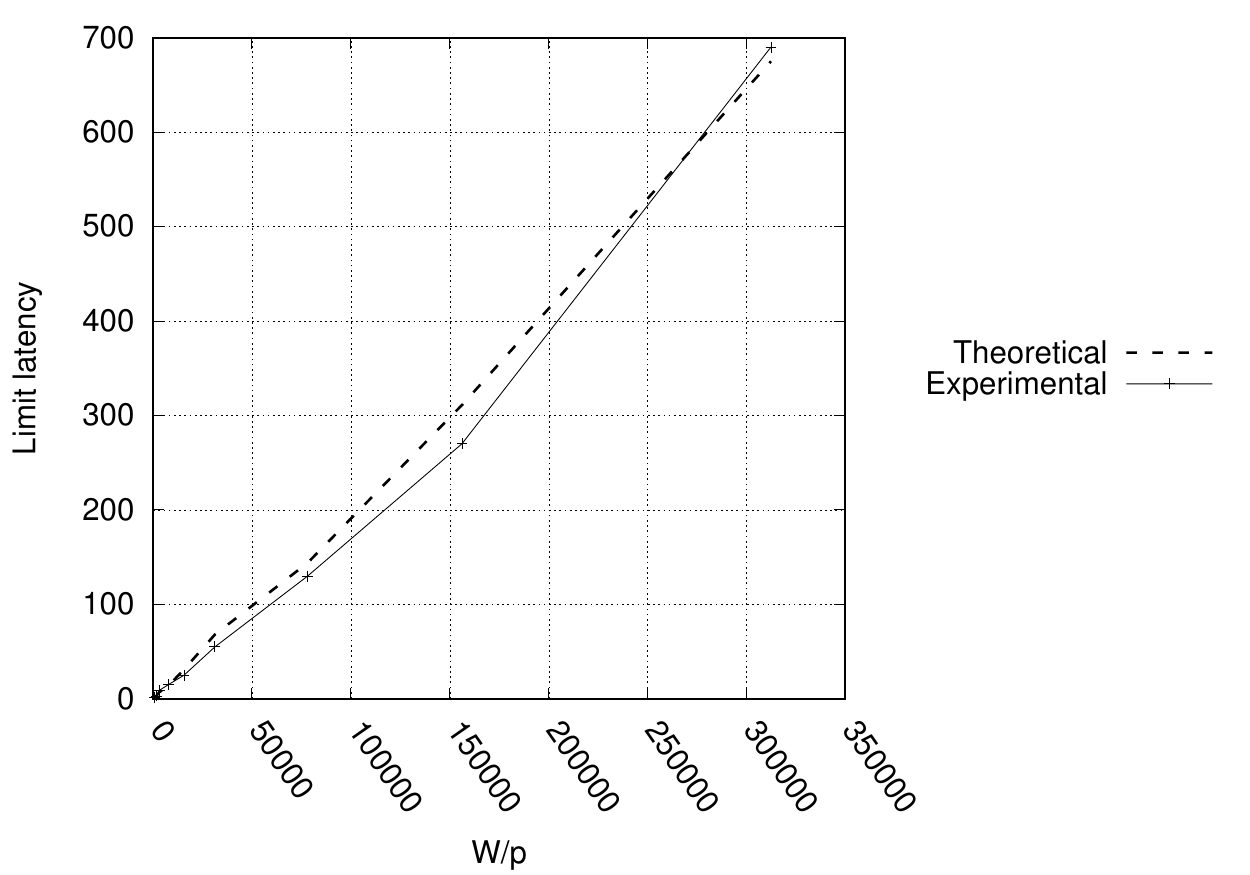}
	\caption{limit latency exhibiting an acceptable Makespan according to $\frac{W}{p}$} 
\label{fig:latency_interval}
\end{figure}
Using this equation we can easily predict when a given $W$,  $p$ and $\lambda$ yields acceptable performance. Moreover for a specific $W$ and a fixed $\lambda$ we can easily choose the maximum number of processors applicable.

To verify the validity of this formula we solve numerically this equation for different $\frac{W}{p}$ to get the theoretical limit latency for an acceptable Makespan. 
We then verify experimentally the obtained solutions. So for a fixed  $W$ and $p$ we test different $\lambda$ and take the maximal one yielding an acceptable Makespan. We call this \textit{the experimental limit latency}. With this result we are able to compare the theoretical and the experimental limit latency.
Fig~\ref{fig:latency_interval} plots the theoretical and experimental limit latency according to $\frac{W}{p}$. The x-axis is $\frac{W}{p}$ for $W$ between $10^5$ and $10^8$ and $p$ between 32 and 256 y-axis show the limit latency.

In Fig~\ref{fig:latency_interval} we observe that the two curves overlap and conclude again on the good accuracy of our prediction. 
Moreover we can see that the relation between the latency limit and the $\frac{W}{p}$
ratio is close to linear. Using this figure we can derive the following equation:
$\frac{W}{p} = 470\lambda$. Using this equation it is easy to evaluate performances for a given $W$, $p$ and $\lambda$. In addition it allows us to compute easily for any configuration the maximal number of processors $\frac{W}{470\lambda}$ allowing an acceptable Makespan.

\subsection{Simultaneous response}
\label{Sec:simultaneousresponse}
We now conclude our experimental campaign by studying the influence of the multiple
work transfers mechanism (\emph{MWT}). In our experimental runs we compare the results
obtained using both variants: With multiple work transfers and with a single work transfer (\emph{SWT}).
We reproduce 1000 runs for $W$ between $10^5$ and $10^8$, $p$ between $8$ and $256$ and $\lambda$ between $2$ and $500$.

The experiments show that the \emph{MWT} mechanism does not bring a significant gain in the overall performances, which spurred us to analyze in detail the execution traces. In this analysis we remark that any execution using a Work Stealing algorithm decomposes into three phases. The first phase which is denoted by the startup phase, when all the processors try to have work. This phase finishes when all processors become active. The second phase corresponds to the situation in which all processors have work and just a few steal requests between processors happen. The last phase starts when there is little work and the majority of processors are inactive.  

\begin{figure}[h!]
	\centering
	\includegraphics[width=\linewidth]{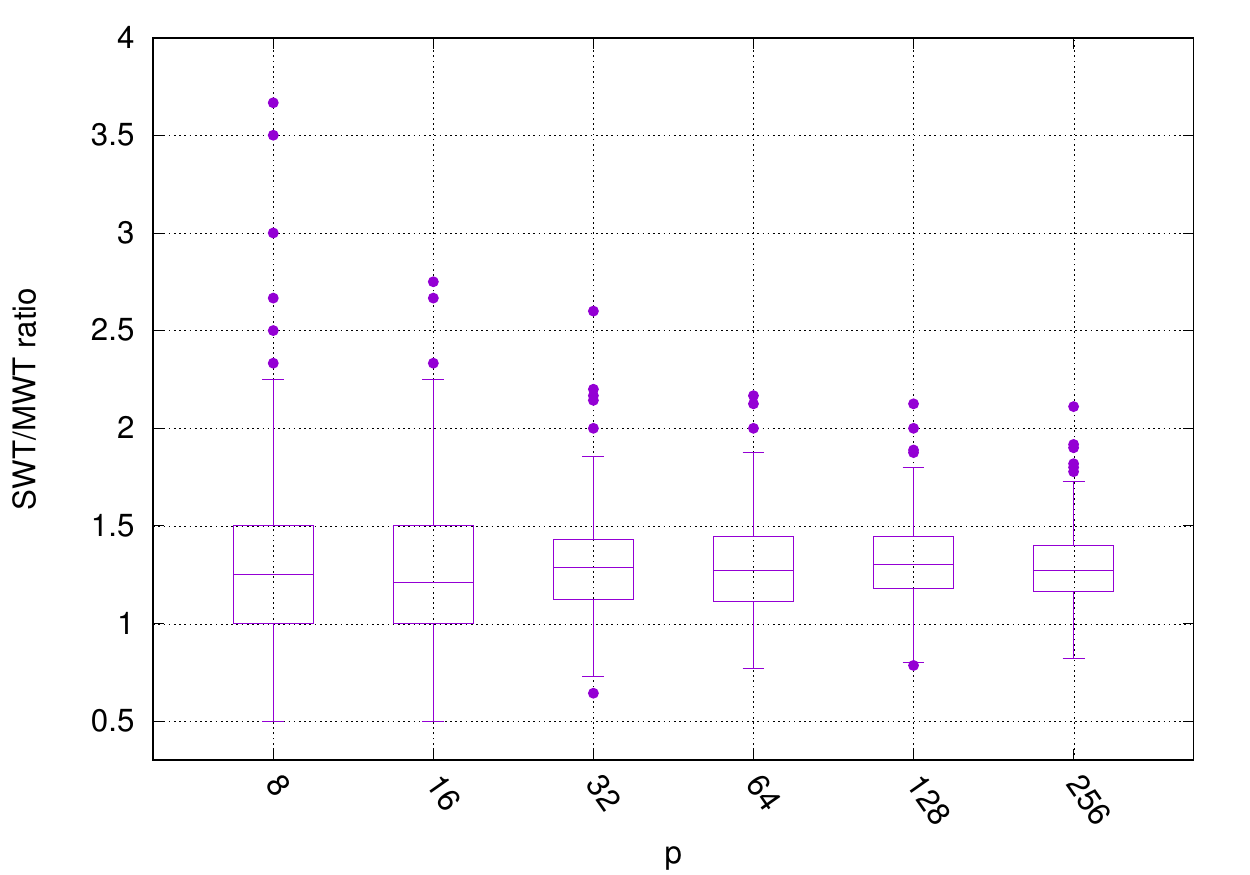}
	\caption{ratio between the duration of the startup phase of the execution using \emph{MWT} and with \emph{SWT} according to the number of processors ($\lambda=250$ and $W=10^8$)} 
\label{fig:simultaneousefig}
\end{figure}

In practice, we observe that the \emph{MWT} mechanism only impacts significantly the startup phase.
Fig~\ref{fig:simultaneousefig} presents in BoxPlot format the ratio between the duration of the startup phase using the \emph{MWT} mechanism and using the \emph{SWT} mechanism according to the number of processors. The x-axis is the number of processors and the y-axis is the ratio between the two durations of the startup phase using the \emph{SWT} and \emph{MWT} mechanisms for $\lambda=250$ and $W=10^8$.  

In this setting we see that \emph{MWT} is reducing the duration of the startup phase for 75\% of the runs with a gain larger than 200\% for a small number of processors.

The behavior of \emph{MWT} is positive on the startup phase but the overall performance gains are small because the duration of the startup phase is small compared to the total execution time. 
We can however imagine different conditions amplifying this effect.
for instance in the case of dependent tasks (for example fork-join graph) would potentially create many startup phases leading to a stronger effect of \emph{MWT}.

\section{Conclusion}
\label{sec:conclusion}
In this paper we study the behavior of the latency on Work Stealing
and its impact on performances.
We base our methodology on observations of execution traces and metrics
under varying combinations of parameters and algorithms.
This work forms the basis of incoming studies on more complex hierarchical topologies.
As such it is important because it allows for a full understanding of the
behavior of different Work Stealing implementations in a base setting.

A first observation on traces is that some useless work transfers
can be avoided by introducing a \emph{steal threshold} limiting the size
of work transfers according to latency. Although this modification does not
improve the execution times in current experiments we believe it will be more useful in a hierarchical
setting where we should be able to set different thresholds based on
distance between senders and receivers.

We also analyze the impact of simultaneous work transfers and show that this
mechanism only impacts the initial distribution of the work. It might therefore
prove itself to be useful under a different tasks model.

The analysis of steal numbers shows they do not grow linearly
with latency. In fact the whole system behaves more like an execution
of Work Stealing in macro steps. This yields an execution time formula
of $\frac{W}{p} + 3.6\log_2(\frac{W}{2\lambda})\lambda$ where the 3.6 constant
is obtained by fitting the formula to simulation results.
The accuracy of this estimation is validated through a large simulation
campaign with an error on average Makespan estimation around 11\%.

This estimation enables us to predict conditions yielding
acceptable performances. Given a latency we can exhibit a relationship
between work and number of processors. Again this estimation will prove
itself to be valuable in future studies on hierarchical platforms because it
indicate us the limits on the minimal amount of work which should be
assigned to each cluster.


%
%



%



\bibliographystyle{IEEEtran}
\bibliography{biblio}

\end{document}